\documentclass[final,1p,times, onecolumn]{elsarticle}
\usepackage{lineno,hyperref,amsmath,amsthm,amssymb,amsfonts,ragged2e,color,subfig}
\journal{``Contributions to Plasma Physics"}
\begin{document}
\begin{frontmatter}
\title{Electrostatic dust-acoustic envelope solitons in an electron depleted plasma}
\author{R.K. Shikha$^{*,1}$, N.A. Chowdhury$^{**,2}$, A. Mannan$^{\ddag1,3}$, and A.A. Mamun$^{\S,1}$}
\address{$^{1}$Department of Physics, Jahangirnagar University, Savar, Dhaka-1342, Bangladesh\\
$^2$ Plasma Physics Division, Atomic Energy Centre, Dhaka-1000, Bangladesh\\
$^3$ Institut f\"{u}r Mathematik, Martin Luther Universit\"{a}t Halle-Wittenberg, Halle, Germany\\
e-mail: $^*$shikha261phy@gmail.com, $^{**}$nurealam1743phy@gmail.com,\\
$^{\ddag}$abdulmannan@juniv.edu, $^{\S}$mamun\_phys@juniv.edu}
\begin{abstract}
A standard nonlinear Schr\"{o}dinger equation has been established by using the reductive perturbation method to investigate the
propagation of electrostatic dust-acoustic waves, and their modulational instability as well as the formation of
localized electrostatic envelope solitons in an electron depleted unmagnetized dusty plasma system comprising opposite polarity
dust grains and super-thermal positive ions. The relevant physical plasma parameters (viz., charge, mass, number
density of positive and negative dust grains, and super-thermality of the positive ions, etc.) have rigorous impact to recognize the stability
conditions of dust-acoustic waves. The present study is useful for understanding the mechanism of the formation of
dust-acoustic envelope solitons associated with dust-acoustic waves in the laboratory and space environments.
\end{abstract}
\end{frontmatter}
\section{Introduction}
The research regarding opposite polarity dusty plasma, which is the combination of
electrons, ions, and highly charged opposite polarity dust grains (DGs), has been increased tremendously due to
their existence in astrophysical environments (viz., asteroid zones \cite{Shukla2001},
interstellar clouds \cite{Shukla2001}, planetary rings \cite{Shukla1992}, Jupiter's magnetosphere \cite{Hossen2016},
cometary tails \cite{Hossen2016}, Earth polar mesosphere \cite{Hossen2016}, and solar system \cite{Hossen2017}, etc.) and
laboratory observation \cite{Barkan1995,Shahmansouri2013}. When energetic
plasma particles (electrons or ions) are incident onto a DG surface, they are either
backscattered/reflected by the DG or they pass through the DG
material. During their passage they may lose their energy partially or fully. A
portion of the lost energy can go into exciting other electrons that in turn may
escape from the material. The emitted electrons are known as secondary electrons.
The release of these secondary electrons from the DG tends to make the
grain surface positive \cite{Shukla2002}. The interaction of photons incident onto the DG
surface causes photoemission of electrons from the DG surface. The DGs, which emit
photoelectrons, may become positively charged \cite{Shukla2002}. The emitted
electrons collide with other DGs and are captured by some of these grains
which may become negatively charged \cite{Shukla2002}. There are, of course, a number of other
DG charging mechanisms, namely thermionic emission, field emission, and impact ionization, etc.

The process by which electrons are inserted to the negatively charged massive DGs from the
background of the dusty plasma medium (DPM) is known as electron depletion, and this electron depleted plasma (EDP)
can be observed in  interstellar clouds \cite{Shukla2001}, cometary tails \cite{Hossen2016},
Earth polar mesosphere \cite{Hossen2016}, Jupiter's magnetosphere \cite{Hossen2016}, solar
system \cite{Hossen2017}, F-rings of Saturn \cite{Sahu2012a}, and laboratory observation \cite{Barkan1995,Shahmansouri2013}.
Shukla and Silin \cite{Shukla1992} considered inertial ions and immobile DGs to
investigate dust-ion-acoustic (DIA) waves (DIAWs) in an EDP. Sahu and Tribeche \cite{Sahu2012a} studied
dust-acoustic (DA) shock waves (DASHWs) and DA solitary waves (DASWs)
in non-planer geometry. Mamun \textit{et al.} \cite{Mamun1996} analyzed electrostatic solitary potential structures in
an EDP, and reported that the existence of large number of ion and DG causes to increase the amplitude of the negative
potentials. Ferdousi \textit{et al.} \cite{Ferdousi2015} examined the DASHWs by considering a two-component EDP,
and demonstrated that under consideration both negative and positive potential structures can exist. Borhanian and Shahmansouri \cite{Borhanian2013}
considered a three-component DPM having inertial massive negative DGs and inertialess two temperature ions, and
investigated DASWs in presence of two temperature super-thermal ions, and highlighted that the phase velocity increases
with ion population but decreases with ion temperature. Mayout and Tribeche \cite{Mayout2012} theoretically analyzed DA double-layers (DADLs)
in an EDP, and graphically recognized that the amplitude of the DADLs causes to decrease with super-thermality of the plasma species.
Sahu and Tribeche \cite{Sahu2012b} studied small amplitude DADLs in a two-component non-thermal EDP.

The parameter $\kappa$ in super-thermal/$\kappa$-distribution can describe the deviation (due to the presence of long range
force fields) of the plasma species from the Maxwellian distribution \cite{Vasyliunas1968,Panwar2014,Eslami2013,Younsi2010,Saini2016}.
The $\kappa$-distribution behaves as Maxwellian distribution for large values of $\kappa$ (i.e., $\kappa\rightarrow\infty$) \cite{Panwar2014,Eslami2013,Younsi2010,Saini2016}. Panwar \textit{et al.} \cite{Panwar2014} demonstrated a theoretical
investigation regarding the propagation of ion-acoustic waves (IAWs) in a three-component plasma having super-thermal electrons,
and found that the amplitude of the compressive (rarefactive) cnoidal
waves increases (decreases) with super-thermality of the electrons. Eslami \textit{et al.} \cite{Eslami2013}
analyzed DIAWs in presence of super-thermal plasma species, and reported that the speed of the DIAWs
increases with the increase in the value of the super-thermality of the plasma species. Younsi and Tribeche \cite{Younsi2010} examined the effects of
excess super-thermal electrons on the formation of electron-acoustic waves (EAWs) in a three-component plasma medium. Saini and Singh \cite{Saini2016}
studied head on collision of two DIAWs in  a multi-component super-thermal plasma, and observed that both amplitude and width of the profile are increasing with $\kappa$.

The nonlinear Schr\"odinger equation (NLSE) is one of the eye-catching equations which can describe the modulational
instability (MI) of various kinds of waves, viz., EAWs \cite{Sultana2011}, IAWs \cite{Gharaee2011}, DIAWs \cite{Jukui2002},
and DA waves (DAWs) \cite{Gill2010}, etc. Sultana and Kourakis \cite{Sultana2011} studied the MI of the EAWs and associated
bright and dark envelope solitons in a three-component super-thermal plasma. Gharaee \textit{et al.} \cite{Gharaee2011}
examined the stability criteria of the IAWs in presence of the super-thermal electrons. Jukui and He \cite{Jukui2002}
theoretically analyzed the MI conditions of the cylindrical and spherical DIAWs. Gill \textit{et al.} \cite{Gill2010}
considered inertial positive and negative DGs and inertialess electrons and ions to study the MI of the DAWs in a multi-component
plasma medium, and reported that the critical wave number which determines the modulationally stable and unstable domains of the
DAWs decreases with increasing the value of $\kappa$.
Borhanian \textit{et al.} \cite{Borhanian2009} numerically examined electromagnetic envelope solitons in magnetized plasma
and found that solitons (bright or dark-type) propagate in the magnetized plasma without any change in amplitude
and shape. In this paper, our aim is to investigate the MI criteria of DAWs and associated envelope solitons in a
three-component EDP having inertial positive and negative DGs and inertialess $\kappa$-distributed ions.

The manuscript is organized as follows: The governing equations are provided
in section \ref{2sec:Governing Equations}. The derivation of the NLSE by using the reductive purturbation method (RPM)
is demonstrated in section \ref{2sec:Derivation of NLSE}.
The MI of DAWs is presented in section \ref{2sec:MI analysis}. The envelope solitons are
presented in section \ref{2sec:Envelope solitons}. The conclusion is shown in section \ref{2sec:Conclusion}.
\section{Governing Equations}
\label{2sec:Governing Equations}
We consider a three-component unmagnetized EDP comprising inertial negatively and positively
charged massive DGs, and $\kappa$-distributed positive ions. At equilibrium, the
quasi-neutrality condition can be written as $Z_in_{i0} + Z_+n_{+0} \approx
Z_-n_{-0}$ ; where $n_{i0}$ is the number densities of positive ions, and
$n_{-0}$ ($n_{+0}$) is the number densities of negative (positive) DGs; $Z_i$ is the charge
state of positive ions, and $Z_-$ ($Z_+$) is the charge state of negative (positive) DGs.
So, the normalizing equations for our plasma model can be written as
\begin{eqnarray}
&&\hspace*{-1.3cm}\frac{\partial n_-}{\partial t}+\frac{\partial}{\partial x}(n_-u_-)=0,
\label{2:eq1}\\
&&\hspace*{-1.3cm}\frac{\partial u_-}{\partial t}+u_-\frac{\partial u_-}{\partial x}+\beta_1n_-\frac{\partial n_{-}}{\partial x} = \frac{\partial\phi}{\partial x},
\label{2:eq2}\\
&&\hspace*{-1.3cm}\frac{\partial n_+}{\partial t}+\frac{\partial}{\partial x}(n_+u_+)=0,
\label{2:eq3}\\
&&\hspace*{-1.3cm}\frac{\partial u_+}{\partial t}+u_+\frac{\partial u_+}{\partial x}+\beta_2n_+\frac{\partial n_{+}}{\partial x} = -\beta_3\frac{\partial\phi}{\partial x},
\label{2:eq4}\\
&&\hspace*{-1.3cm}\frac{\partial^{2} \phi}{\partial x^{2}}=n_--(1-\beta_4)n_i-\beta_4n_+,
\label{2:eq5}\
\end{eqnarray}
where $n_i$, $n_-$, and $n_+$ are normalized by $n_{i0}$, $n_{-0}$, and $n_{+0}$, respectively; $u_+$ ($u_-$)
represents the positive (negative) dust fluid speed which is normalized by the DA wave speed $C_{-}=(Z_-k_BT_i/m_-)^{1/2}$
(with $T_i$ being temperature of ion, $m_+$ being positive dust mass, and $k_B$ being the Boltzmann constant);
$\phi$ represents the electrostatic wave potential normalized by $k_BT_i/e$ (with $e$ being the magnitude of single
electron charge); the time and space variables are, respectively, normalized by  $\omega_{P_-}^{-1}=(m_-/4\pi e^{2}Z_-^{2}n_{-0})^{1/2}$
and $\lambda_{D_-} = (k_BT_i/4\pi e^{2}Z_-n_{-0})^{1/2}$. $P_-= P_{-0} (N_-/n_{-0})^{\gamma}$ (with $P_{-0}$ being
the equilibrium pressure term of the negative DGs), $P_+= P_{+0} (N_+/n_{+0})^{\gamma}$ (with $P_{+0}$ being
the equilibrium pressure term of the positive DGs), and $\gamma=(N + 2)/N$, where $N$ is the degree of freedom and
for one-dimensional case $N=1$, then $\gamma=3$; $P_{-0} = n_{-0} k_B T_-$, $P_{+0}=n_{+0} k_B T_+$
(with $T_-$ and $T_+$ being the temperature of negative and positive DGs);
and other parameters are $\beta_1=3T_-/Z_-T_i$, $\beta_2=3T_+m_-/Z_-T_im_+$, $\beta_3=Z_+m_-/Z_-m_+$, and
$\beta_4=Z_+n_{+0}/Z_-n_{-0}$. It may be noted here that we have considered $m_->m_+$, $Z_->Z_+$, and $n_{-0}>n_{+0}$.
The expression for the number density of ions following the $\kappa$-distribution \cite{Shahmansouri2013} can be written as
\begin{eqnarray}
&&\hspace*{-1.3cm}n_i=\left[1 +\frac{\phi}{(\kappa-3/2)}\right]^{-\kappa+\frac{1}{2}}
\label{2:eq6}\
\end{eqnarray}
where the parameter $\kappa$ is known as super-thermality of the positive ions.
Now, by substituting Eq. \eqref{2:eq6} into Eq. \eqref{2:eq5}, and expanding the term $\phi$ up to
third order, we obtain
\begin{eqnarray}
&&\hspace*{-1.3cm}\frac{\partial^{2} \phi}{\partial x^{2}}+\beta_4n_+=(\beta_4-1)+n_-+G_1\phi+G_2\phi^{2}+G_3\phi^{3}+\cdot\cdot\cdot,
\label{2:eq7}\
\end{eqnarray}
where
\begin{eqnarray}
&&\hspace*{-1.3cm}G_1=\frac{(1-\beta_4)(2\kappa-1)}{(2\kappa-3)},~~~~~G_2 = -\frac{(1-\beta_4)(2\kappa-1)(2\kappa+1)}{2(2\kappa-3)^2},
\nonumber\\
&&\hspace*{-1.3cm}G_3=\frac{(1-\beta_4)(2\kappa-1)(2\kappa+1)(2\kappa+3)}{6(2\kappa-3)^{3}}.
\nonumber\
\end{eqnarray}
\section{Derivation of the NLSE}
\label{2sec:Derivation of NLSE}
To study the MI of DAWs, we will derive the NLSE by employing
the RPM. So, we first introduce
the stretched co-ordinates \cite{C1,C2,C3,C4,C5,C6}
\begin{eqnarray}
&&\hspace*{-1.3cm}\xi={\epsilon}(x-v_g t),
\label{2:eq8}\\
&&\hspace*{-1.3cm}\tau={\epsilon}^{2} t,
\label{2:eq9}\
\end{eqnarray}
where $v_g$ is the group speed and $\epsilon$ is a small parameter. Then, we can write the dependent variables as
\begin{eqnarray}
&&\hspace*{-1.3cm}n_-=1+\sum_{m=1}^{\infty}\epsilon^{m}\sum_{l=-\infty}^{\infty}n_{-l}^{(m)}(\xi,\tau)\mbox{exp}[i l(kx-\omega t)],
\label{2:eq10}\\
&&\hspace*{-1.3cm}u_-=\sum_{m=1}^{\infty}\epsilon^{m}\sum_{l=-\infty}^{\infty}u_{-l}^{(m)}(\xi,\tau)\mbox{exp}[i l(kx-\omega t)],
\label{2:eq11}\\
&&\hspace*{-1.3cm}n_+=1+\sum_{m=1}^{\infty}\epsilon^{m}\sum_{l=-\infty}^{\infty}n_{+l}^{(m)}(\xi,\tau)\mbox{exp}[i l(kx-\omega t)],
\label{2:eq12}\\
&&\hspace*{-1.3cm}u_+=\sum_{m=1}^{\infty}\epsilon^{m}\sum_{l=-\infty}^{\infty}u_{+l}^{(m)}(\xi,\tau)\mbox{exp}[i l(kx-\omega t)],
\label{2:eq13}\\
&&\hspace*{-1.3cm}\phi=\sum_{m=1}^{\infty}\epsilon^{m}\sum_{l=-\infty}^{\infty}\phi_l^{(m)}(\xi,\tau)\mbox{exp}[i l(kx-\omega t)],
\label{2:eq14}\
\end{eqnarray}
where $k$ ($\omega$) is real variable representing the carrier wave number (frequency).
The derivative operators in the above equations are treated as follows \cite{C7,C8,C9,C10,C11,C12,C13}:
\begin{eqnarray}
&&\hspace*{-1.3cm}\frac{\partial}{\partial t}\rightarrow\frac{\partial}{\partial t}-\epsilon v_g\frac{\partial}{\partial\xi}+\epsilon^2\frac{\partial}{\partial\tau},
\label{2:eq15}\\
&&\hspace*{-1.3cm}\frac{\partial}{\partial x}\rightarrow\frac{\partial}{\partial x}+\epsilon\frac{\partial}{\partial\xi}.
\label{2:eq16}\
\end{eqnarray}
Now, by substituting Eqs. \eqref{2:eq10}$-$\eqref{2:eq16} into Eqs. \eqref{2:eq1}$-$\eqref{2:eq4} and Eq.
\eqref{2:eq7}, and collecting the terms containing $\epsilon$, the first order ($m=1$ with $l=1$) equations
can be expressed as
\begin{eqnarray}
&&\hspace*{-1.3cm}\omega n_{-1}^{(1)}=ku_{-1}^{(1)},
\label{2:eq17}\\
&&\hspace*{-1.3cm}k\phi_1^{(1)}=k\beta_1n_{-1}^{(1)}-\omega u_{-1}^{(1)},
\label{2:eq18}\\
&&\hspace*{-1.3cm}\omega n_{+1}^{(1)}=ku_{+1}^{(1)},
\label{2:eq19}\\
&&\hspace*{-1.3cm}k\beta_3\phi_1^{(1)}=\omega u_{+1}^{(1)}-k\beta_2n_{+1}^{(1)},
\label{2:eq20}\\
&&\hspace*{-1.3cm}n_{-1}^{(1)}=\beta_4n_{+1}^{(1)}-k^{2}\phi_1^{(1)}-G_1\phi_1^{(1)},
\label{2:eq21}\
\end{eqnarray}
these equations reduce to
\begin{eqnarray}
&&\hspace*{-1.3cm}n_{-1}^{(1)}=\frac{k^2}{M}\phi_1^{(1)},
\label{2:eq22}\\
&&\hspace*{-1.3cm}u_{-1}^{(1)}=\frac{\omega k}{M}\phi_1^{(1)},
\label{2:eq23}\\
&&\hspace*{-1.3cm}n_{+1}^{(1)}=\frac{\beta_3k^2}{N}\phi_1^{(1)},
\label{2:eq24}\\
&&\hspace*{-1.3cm}u_{+1}^{(1)}=\frac{\omega k\beta_3}{N}\phi_1^{(1)},
\label{2:eq25}\
\end{eqnarray}
where $M=\beta_1k^{2}-\omega^{2}$ and $N=\omega^{2}-\beta_2k^{2}$.
We thus obtain the dispersion relation of DAWs
\begin{eqnarray}
&&\hspace*{-1.3cm}\omega^2=\frac{k^{2}T \pm k^{2}\sqrt{T^{2}-4(G_1+k^{2})L}}{2(G_1+k^{2})},
\label{2:eq26}\
\end{eqnarray}
where $L=\beta_2+\beta_1\beta_2G_1+\beta_1\beta_3\beta_4+\beta_1\beta_2k^{2}$ and $T=1+\beta_1G_1+\beta_2G_1+\beta_3\beta_4+\beta_1k^{2}+\beta_2k^{2}$.
In Eq. \eqref{2:eq26}, to get real and positive values of $\omega$, the condition $T^2>4(G_1+k^{2})L$ should be
satisfied. The positive and negative signs in Eq. \eqref{2:eq26} corresponds to the fast ($\omega_f$) and slow ($\omega_s$) DA modes.
The fast DA mode corresponds to the case in which both inertial DGs oscillate in phase with the inertialess
ions. On the other hand, the slow DA mode corresponds to the case in which only one of the inertial DGs
oscillates in phase with inertialess ions, but the other inertial DG in anti-phase with
them \cite{Dubinov2009,Saberiana2017}.
The second-order ($m=2$ with $l=1$) equations are given by
\begin{eqnarray}
&&\hspace*{-1.3cm}n_{-1}^{(2)}=\frac{k^{2}}{M}\phi_1^{(2)}+\frac{i(k\omega^{2}-2v_g\omega k^{2}-kM+\beta_1k^{3})}{M^{2}}\frac{\partial \phi_1^{(1)}}{\partial\xi},
\label{2:eq27}\\
&&\hspace*{-1.3cm}u_{-1}^{(2)}=\frac{k\omega}{M}\phi_1^{(2)}+\frac{i(\omega^{3}+\beta_1\omega k^{2}-2v_gk\omega^{2}-v_gkM)}{M^{2}}\frac{\partial \phi_1^{(1)}}{\partial\xi},
\label{2:eq28}\\
&&\hspace*{-1.3cm}n_{+1}^{(2)}=\frac{\beta_3k^{2}}{N}\phi_1^{(2)}-\frac{i\beta_3(\beta_2k^{3}-2\omega v_gk^{2}+k{\omega^{2}+kN})}{N^{2}}\frac{\partial \phi_1^{(1)}}{\partial\xi},
\label{2:eq29}\\
&&\hspace*{-1.3cm}u_{+1}^{(2)}=\frac{\beta_3k\omega}{N}\phi_1^{(2)}-\frac{i\beta_3(\beta_2\omega k^{2}-2v_gk\omega^{2}+v_gkN+\omega^{3})}{N^{2}}\frac{\partial \phi_1^{(1)}}{\partial\xi},~~~~~~~~
\label{2:eq30}\
\end{eqnarray}
with compatibility condition, we can find the group velocity
\begin{eqnarray}
&&\hspace*{-1.3cm}v_g=\frac{k^{2}(\beta_1N^{2}+\beta_2\beta_3\beta_4M^{2})+\omega^{2}(N^{2}+\beta_3\beta_4M^{2})-2M^{2}N^{2}-MN(N-\beta_3\beta_4M)}{2\omega k(N^{2}+\beta_3\beta_4M^{2})}.
\label{2eq:31}\
\end{eqnarray}
The coefficients of $\epsilon$ for $m=2$ with $l=2$ provide the second order
harmonic amplitudes which are found to be proportional
to $|\phi_1^{(1)}|^{2}$
\begin{eqnarray}
&&\hspace*{-1.3cm}n_{-2}^{(2)}=G_4|\phi_1^{(1)}|^{2},
\label{2eq:32}\\
&&\hspace*{-1.3cm}u_{-2}^{(2)}=G_5 |\phi_1^{(1)}|^{2},
\label{2eq:33}\\
&&\hspace*{-1.3cm}n_{+2}^{(2)}=G_6|\phi_1^{(1)}|^{2},
\label{2eq:34}\\
&&\hspace*{-1.3cm}u_{+2}^{(2)}=G_7 |\phi_1^{(1)}|^{2},
\label{2eq:35}\\
&&\hspace*{-1.3cm}\phi_{2}^{(2)}=G_8 |\phi_1^{(1)}|^{2},
\label{2eq:36}\
\end{eqnarray}
where
\begin{eqnarray}
&&\hspace*{-1.1cm}G_4=\frac{2G_8k^{2}M^{2}-(3\omega^{2}k^{4}+\beta_1k^{6})}{2M^{3}},
\nonumber\\
&&\hspace*{-1.1cm}G_5=\frac{\omega G_4M^{2}-\omega k^{4}}{kM^{2}},
\nonumber\\
&&\hspace*{-1.1cm}G_6=\frac{2\beta_3G_8k^{2}N^{2}+3\omega^{2}\beta_3^{2}k^{4}+\beta_2\beta_3^{2}k^{6}}{2N^{3}},
\nonumber\\
&&\hspace*{-1.1cm}G_7=\frac{\omega G_6N^{2}-\omega\beta_3^{2}k^{4}}{kN^{2}},
\nonumber\\
&&\hspace*{-1.1cm}G_8=\frac{N^{3}(3\omega^{2}k^{4}+\beta_1k^{6})
+\beta_4M^{3}(3\omega^{2}\beta_3^{2}k^{4}+\beta_2\beta_3^{2}k^{6})
-2G_2M^{3}N^{3}}{2M^{2}k^{2}N^{3}+2M^{3}N^{3}(4k^{2}+G_1)-2\beta_3\beta_4N^{2}k^{2}M^{3}}.
\nonumber\
\end{eqnarray}
Now, we consider the expression for ($m=3$ with $l=0$) and
($m=2$ with $l=0$) which leads to the zeroth harmonic modes.
Thus, we obtain
\begin{eqnarray}
&&\hspace*{-1.3cm}n_{-0}^{(2)}=G_{9}|\phi_1^{(1)}|^{2},
\label{2:eq37}\\
&&\hspace*{-1.3cm}u_{-0}^{(2)}=G_{10}|\phi_1^{(1)}|^{2},
\label{2:eq38}\\
&&\hspace*{-1.3cm}n_{+0}^{(2)}=G_{11}|\phi_1^{(1)}|^{2},
\label{2:eq39}\\
&&\hspace*{-1.3cm}u_{+0}^{(2)}=G_{12}|\phi_1^{(1)}|^{2},
\label{2:eq40}\\
&&\hspace*{-1.3cm}\phi_0^{(2)}=G_{13}|\phi_1^{(1)}|^{2},
\label{2:eq41}\
\end{eqnarray}
where
\begin{eqnarray}
&&\hspace*{1.0cm}G_9=\frac{\beta_1k^{4}+\omega^{2}k^{2}+2v_g\omega k^{3}-G_{13}M^{2}}{M^{2}(v_g^{2}-\beta_1)},
\nonumber\\
&&\hspace*{1.0cm}G_{10}=\frac{v_gG_9M^{2}-2\omega k^{3}}{M^{2}},
\nonumber\\
&&\hspace*{1.0cm}G_{11}=\frac{2v_g\omega \beta_3^{2}k^{3}+\omega^{2}k^{2}\beta_3^{2}+\beta_2\beta_3^{2}k^{4}+\beta_3G_{13}N^{2}}{N^{2}(v_g^{2}-\beta_2)},
\nonumber\\
&&\hspace*{1.0cm}G_{12}=\frac{v_gG_{11}N^{2}-2\omega\beta_3^{2}k^{3}}{N^{2}},
\nonumber\\
&&\hspace*{1.0cm}G_{13}=\frac{N^{2}(2v_g\omega k^{3}+\beta_1k^{4}+k^{2}\omega^{2})(v_g^{2}-\beta_2)+2G_2M^{2}N^{2}(v_g^{2}-\beta_1)(v_g^{2}-\beta_2)
-\mathcal{S}_1}{\beta_3\beta_4M^{2}N^{2}(v_g^{2}-\beta_1)+M^{2}N^{2}(v_g^{2}-\beta_2)
-\mathcal{S}_2}.
\nonumber\
\end{eqnarray}
where $\mathcal{S}_1=\beta_4M^{2}(2v_g\omega\beta_3^{2}k^{3}+\beta_2\beta_3^{2}k^4+\beta_3^{2}k^{2}\omega^{2})(v_g^{2}-\beta_1)$
and $\mathcal{S}_2=G_1M^{2}N^{2}(v_g^{2}-\beta_1)(v_g^{2}-\beta_2)$. Finally, the third harmonic modes ($m=3$) and ($l=1$), with the
help of Eqs. \eqref{2:eq22}$-$\eqref{2:eq41}, give a set of equations which can be reduced
to the following NLSE:
\begin{eqnarray}
&&\hspace*{-1.3cm}i \frac{\partial\Phi}{\partial\tau} + P \frac{\partial^{2}\Phi}{\partial\xi^{2}}+ Q|\Phi|^{2}\Phi =0,
\label{2:eq42}\
\end{eqnarray}
where $\Phi=\phi_1^{(1)}$ is used for simplicity. The dispersion coefficient
$P$ is
\begin{eqnarray}
&&\hspace*{-0.7cm}P=\frac{N^{3}\{(v_g\omega-\beta_1k)(\beta_1k^{3}-2v_g\omega k^{2}+k\omega^{2}-kM)+(v_gk-\omega)(\beta_1\omega k^{2}-2v_gk\omega^{2}+\omega^{3}-v_gkM)\}-\mathcal{S}_3}{2\omega k^{2}MN(N^{2}+\beta_3\beta_4M^{2})},
\nonumber\
\end{eqnarray}
where $\mathcal{S}_3=M^{3}N^{3}+\beta_3\beta_4M^{3}\{(v_gk-\omega)(\beta_2\omega k^{2}-2v_gk\omega^{2}+\omega^{3}+v_gkN)
-(v_g\omega-\beta_2k)(\beta_2k^{3}-2\omega v_gk^{2}+k\omega^{2}+kN)\}$ and the nonlinear coefficient $Q$ is
\begin{eqnarray}
&&\hspace*{-0.7cm}Q=\frac{2G_2M^{2}N^{2}(G_8+G_{13})+3G_3M^{2}N^{2}-2\omega N^{2}k^{3}(G_5+G_{10})-2\omega\beta_3\beta_4 M^{2}k^{3}(G_7+G_{12})-\mathcal{S}_4}{2\omega k^{2}(N^{2}+\beta_3\beta_4M^{2})},
\nonumber\
\end{eqnarray}
where $\mathcal{S}_4=N^{2}(\omega^{2}k^{2}+\beta_1 k^{4})(G_4+G_9)+M^{2}(\beta_3\beta_4\omega^{2}k^{2}+\beta_2\beta_3\beta_4k^{4})(G_6+G_{11})$.
It may be noted here that both $P$ and $Q$ are functions of various
plasma parameters such as $k$, $\beta_3$, $\beta_4$, and $\kappa$. So, all the plasma
parameters are used to maintain the nonlinearity and the dispersion
properties of the EDP.
\begin{figure}[t!]
\centering
\includegraphics[width=80mm]{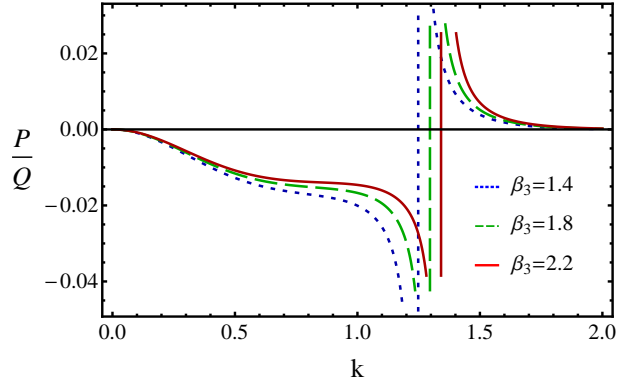}
\caption{Plot of $P/Q$ vs $k$ for different values of $\beta_3$ when $\beta_1=0.006$, $\beta_2=0.06$, $\beta_4=0.8$, $\kappa=1.7$, and $\omega_f$.}
\label{2Fig:F1}
\end{figure}
\begin{figure}[t!]
\centering
\includegraphics[width=80mm]{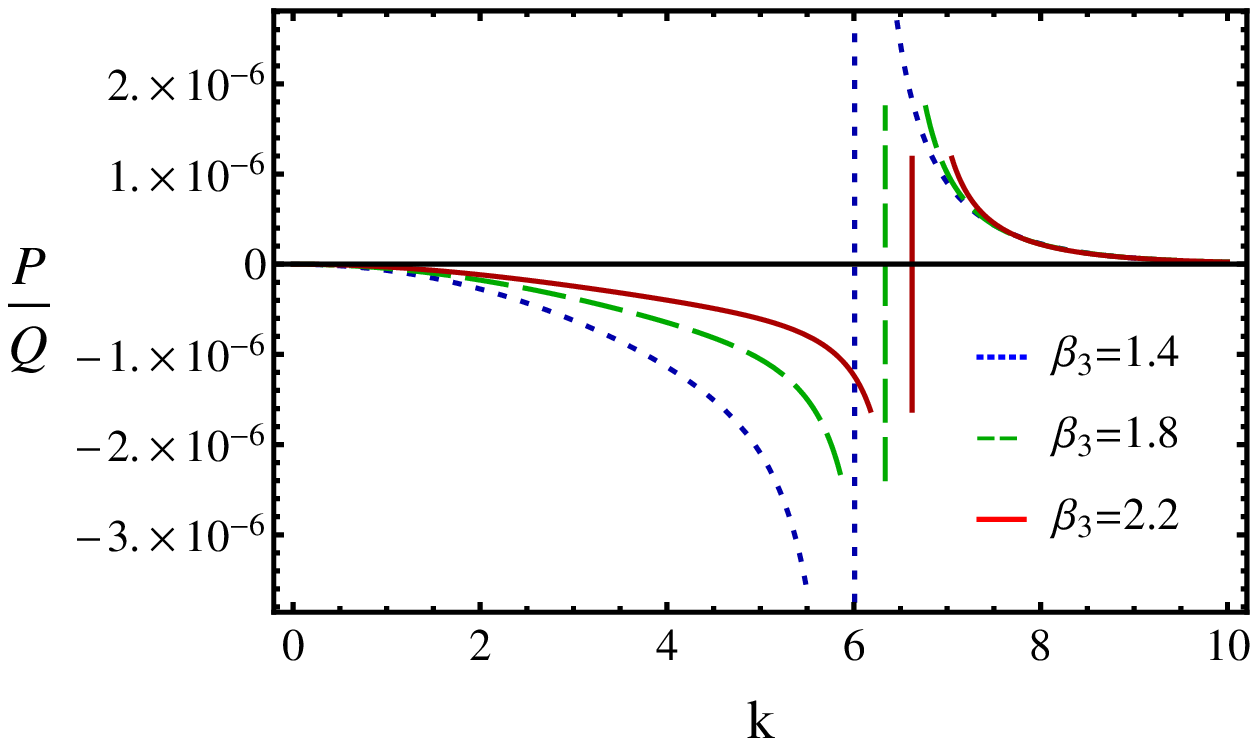}
\caption{Plot of $P/Q$ vs $k$ for different values of $\beta_3$  when $\beta_1=0.006$, $\beta_2=0.06$, $\beta_4=0.8$, $\kappa=1.7$, and $\omega_s$.}
 \label{2Fig:F2}
\end{figure}
\begin{figure}[t!]
\centering
\includegraphics[width=80mm]{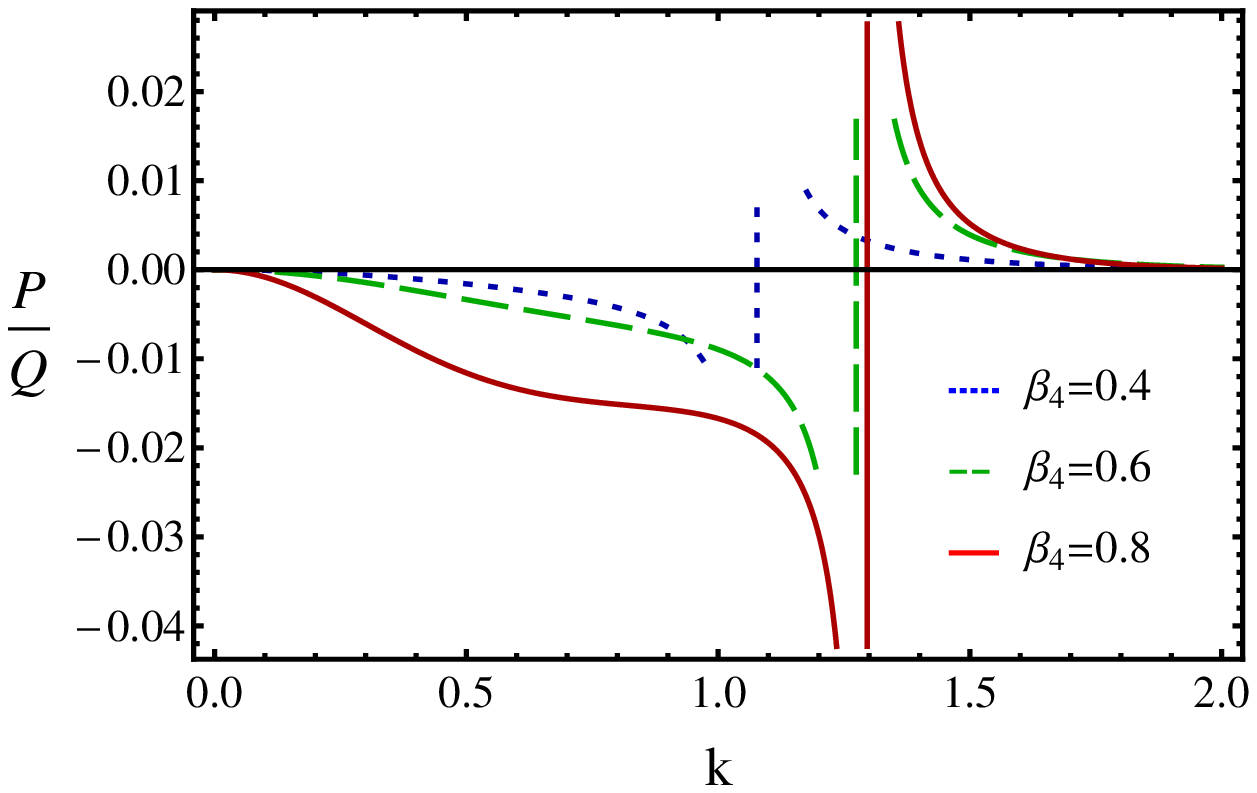}
\caption{Plot of $P/Q$ vs $k$ for different values of $\beta_4$ when $\beta_1=0.006$, $\beta_2=0.06$, $\beta_3=1.8$, $\kappa=1.7$, and $\omega_f$.}
\label{2Fig:F3}
\vspace{0.8cm}
\includegraphics[width=80mm]{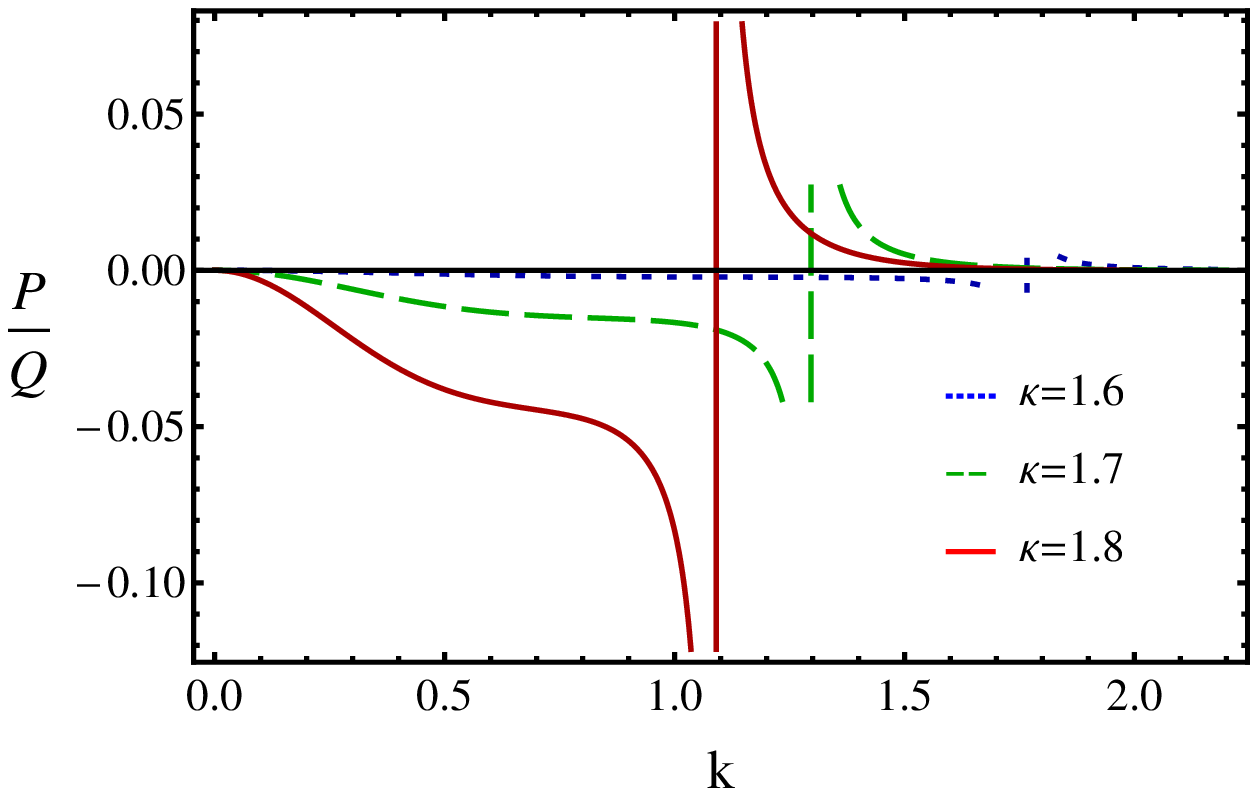}
\caption{Plot of $P/Q$ vs $k$ for different values of $\kappa$ when $\beta_1=0.006$, $\beta_2=0.06$, $\beta_3=1.8$, $\beta_4=0.8$, and $\omega_f$.}
 \label{2Fig:F4}
\end{figure}
\begin{figure}[t!]
\centering
\includegraphics[width=80mm]{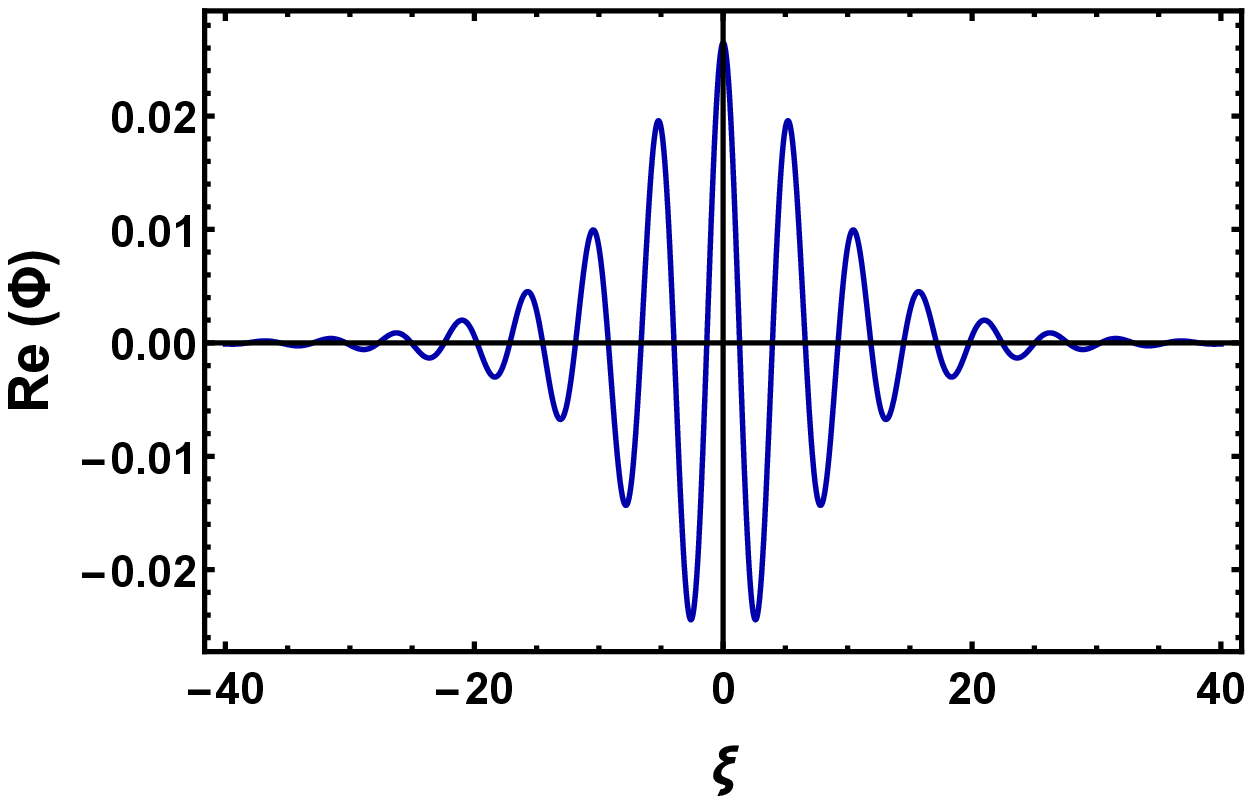}
\caption{The variation of $Re(\Phi)$ vs $\xi$ for bright envelope solitons when $k=1.4$, $\tau=0$, $\beta_1=0.006$, $\beta_2=0.06$, $\beta_3=1.8$, $\beta_4=0.8$, $\psi_0=0.0007$, $\tau=0$, $U=0.5$, $\Omega_0=0.2$, $\kappa=1.7$, and $\omega_f$.}
\label{2Fig:F5}
\end{figure}
\vspace{0.8cm}
\begin{figure}[t!]
\centering
\includegraphics[width=80mm]{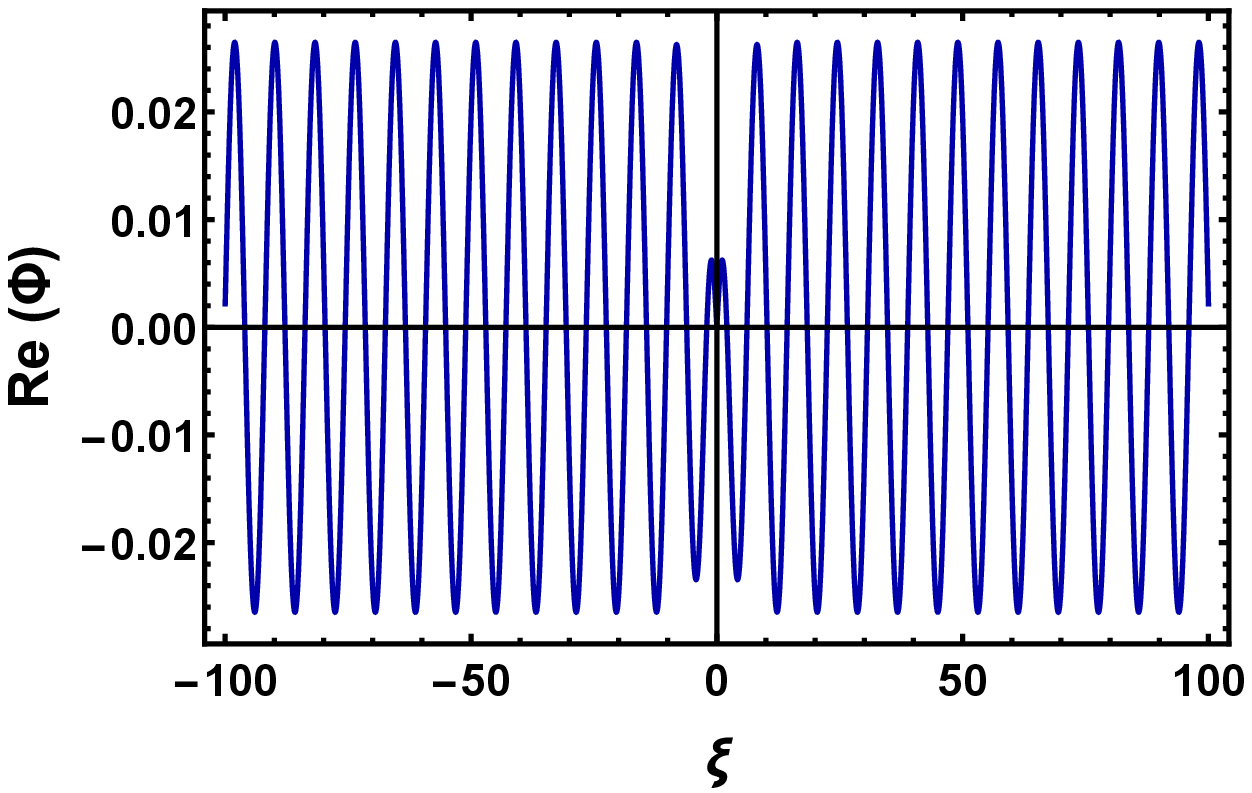}
\caption{The variation of $Re(\Phi)$ vs $\xi$ for dark envelope solitons when $k=0.2$, $\tau=0$, $\beta_1=0.006$, $\beta_2=0.06$, $\beta_3=1.8$, $\beta_4=0.8$, $\psi_0=0.0007$, $\tau=0$, $U=0.5$, $\Omega_0=0.2$, $\kappa=1.7$, and $\omega_f$.}
\label{2Fig:F6}
\end{figure}
\section{Modulational instability}
\label{2sec:MI analysis}
The space and time evolution of the DAWs in an EDP medium are directly governed by the  dispersion
($P$) and nonlinear  ($Q$) coefficients of NLSE, and are indirectly governed by different plasma parameters such as
as $k$, $\beta_3$, $\beta_4$, and $\kappa$. Thus, these plasma parameters significantly
change the stability conditions of DAWs. The stable and unstable parametric regimes of DAWs are
organised by the sign of $P$ and $Q$ of Eq. \eqref{2:eq42} \cite{Sultana2011,Gharaee2011,Jukui2002,Gill2010}.
When $P$ and $Q$ have the same sign (i.e., $P/Q > 0$), the evolution of DAWs amplitude is modulationally
unstable in the presence of external perturbations, and allows to generate bright envelope solitons.
On the other hand, when $P$ and $Q$ have opposite signs (i.e., $P/Q < 0$), the evolution of DAWs amplitude
is modulationally stable in the presence of external perturbations, and allows to generate dark envelope solitons.
The plot of $P/Q$ against $k$ yields stable and unstable parametric regimes of the DAWs. The point, at which the
transition of $P/Q$ curve intersects with the $k$-axis, is known as the threshold
or critical wave number $k~(= k_c)$ \cite{Sultana2011,Gharaee2011,Jukui2002,Gill2010}.

We have investigated the stable/unstable parametric regimes for the DAWs by depicting $P/Q$ versus
$k$ graph for different values of $\beta_3$ in Fig. \ref{2Fig:F1} (under the consideration of fast
mode) and in Fig. \ref{2Fig:F2} (under the consideration of slow mode).
It is clear from these figures that (a) for both fast and slow modes, DAWs are modulationally stable (i.e., $P$ and $Q$ have opposite sign)
and unstable (i.e., $P$ and $Q$ have same sign) for small values of $k$;
(b) the $k_c$ increases with an increase in the value of $\beta_3$;
(c) the charge state of the negative dust ($Z_-$) reduces the $k_c$ as well as destabilize the DAWs for small values
of $k$ while the charge state of the positive dust ($Z_+$) increases the $k_c$ as well as destabilize the DAWs for large
values of $k$ when their masses remain constant; (d) in fast mode, DAWs are modulationally unstable for small value of $k$ ($k\cong1.2$)
while in slow mode, DAWs are modulationally unstable for large value of $k$ ( $k\cong6.0$) with respect to the
fast mode when other plasma parameters remain constant.

Figure \ref{2Fig:F3} describes the effects of the number density of the positive and negative dust grains and their charge state in
recognizing the stable and unstable regions of the DAWs. It is clear from this figure that (a) as we increase $\beta_4$, the $k_c$
increases as well as destabilize the DAWs for large values of $k$; (b) the increase in the value of the positive (negative) dust grains
number density causes to increase (decrease) the $k_c$ for a constant value of positive and negative dust grains charge state (via $\beta_4$).
The super-thermal ions of EDP can easily demonstrate the stability criterion of the DAWs, and it is obvious from Fig. \ref{2Fig:F4} that as we increase the value of $\kappa$, the $k_c$ decreases as well as destabilize the DAWs for small values of $k$.
\section{Envelope solitons}
\label{2sec:Envelope solitons}
The envelope solitonic solutions of the NLSE \eqref{2:eq42}, which
can be obtained by a number of straightforward mathematical
steps, are available in a large number of existing literature \cite{Sultana2011}.
The bright envelope solitons corresponding to the unstable parametric regime (i.e., $P/Q > 0$) can be written as
\begin{eqnarray}
&&\hspace*{-1.3cm}\Phi(\xi,\tau)=\left[\psi_0~\mbox{sech}^{2} \left(\frac{\xi-U\tau}{W}\right)\right]^{1/2}
\times~\exp \left[\frac{i}{2P}\left\{U\xi+\left(\Omega_0-\frac{U^{2}}{2}\right)\tau \right\}\right],
\label{2eq:43}\
\end{eqnarray}
where $\psi_0$ is the amplitude of localized pulse for both bright and dark
envelope soliton, $U$ is the propagation speed of the localized pulse, $W$ is the soliton width, and
$\Omega_0$ is the oscillating frequency at $U=0$. The soliton width $W$ and the maximum amplitude  $\psi_0$
are related by $W=\sqrt{2\mid P/Q\mid/\psi_0}$. We have exhibited the bright envelope solitons in Fig. \ref{2Fig:F5}.
On the other hand, the dark envelope solitons corresponding
to the stable parametric regime (i.e., $P/Q < 0$) can be written as
\begin{eqnarray}
&&\hspace*{-1.3cm}\Phi(\xi,\tau)=\left[\psi_0~\mbox{tanh}^{2} \left(\frac{\xi-U\tau}{W}\right)\right]^{1/2}
\times~\exp \left[\frac{i}{2P}\left\{U\xi-\left(\frac{U^{2}}{2}-2 P Q \psi_0\right)\tau \right\}\right].
\label{2eq:44}\
\end{eqnarray}
We have exhibited the dark envelope solitons in Fig. \ref{2Fig:F6}.
\section{Conclusion}
\label{2sec:Conclusion}
In this paper, we have theoretically and numerically analysed the criteria of MI of DAWs and
associated bright and dark envelope solitons in a three-component EDP having inertial positive
and negative DGs and inertialess super-thermal electrons. We have employed RPM
for deriving the NLSE. The EDP medium under consideration
supports the stable and unstable DAWs depending on the sign of the ratio of $P$ and $Q$.
The relevant physical plasma parameters (viz., charge, mass, number
density of positive and negative dust grains, and super-thermality of the ions) play an important role
in recognizing the stability conditions as well as generation of the bright and dark electrostatic envelope
solitons. It may be noted here that the effects of gravitational and the magnetic fields are very important
but beyond the scope of our present work. In future and for
better understanding, someone can investigate the nonlinear
propagation in a  three-component EDP by considering the effects of these
gravitational and magnetic fields.
The findings of our present investigation should be useful to understand the nonlinear phenomena (viz. MI and envelope
solitons) in space plasma (i.e., interstellar clouds \cite{Shukla2001}, cometary tails \cite{Hossen2016}, and F-rings of Saturn \cite{Sahu2012a}, etc.)
and laboratory experiments.
\section*{acknowledgements}
The authors are grateful to the anonymous reviewer
for his/her constructive suggestions which have significantly
improved the quality of our manuscript. A. Mannan thanks the Alexander von Humboldt Foundation for a Postdoctoral Fellowship.

\end{document}